\newif\ifarxiv
\arxivtrue

\documentclass[sigconf,nonacm]{acmart}

\usepackage{xcolor,soul,multicol}
\usepackage{listings}
\lstdefinestyle{CStyle} {
basicstyle=\ttfamily\scriptsize,       
language=C,
keywordstyle=\color{blue}\ttfamily\bfseries,
keywordstyle=[1]{\color{blue}\ttfamily\bfseries},
keywordstyle=[2]{\color{blue}\ttfamily\bfseries},
stepnumber=1,                   
numbersep=8pt,                  
breakindent=0pt,
firstnumber=1,
showspaces=false,               
showstringspaces=false,         
showtabs=false,                 
tabsize=2,  		
captionpos=b,   		
breaklines=false,    	
breakatwhitespace=true,    
columns=fixed,
basewidth=0.52em,
numberblanklines=false,
escapechar=|,
morekeywords={diff, query, operation, merge, state, and}
}
\usepackage{graphicx}
\graphicspath{ {./images/} }
\usepackage{caption}
\usepackage{subcaption}

\newcommand{\smallitem}[1]{\vspace{0.3em}\noindent\textbf{#1}}
\newcommand{\smallitembot}{\vspace{0.5em}\noindent}

\newif\ifcomments
\ifcomments
    \providecommand{\shadaj}[1]{{\protect\color{brown}{\bf [shadaj: #1]}}}
    \providecommand{\conor}[1]{{\protect\color{red}{\bf [conor: #1]}}}
    \providecommand{\alvin}[1]{{\protect\color{purple}{\bf [alvin: #1]}}}
    \providecommand{\mae}[1]{{\protect\color{blue}{\bf [mae: #1]}}}
    \providecommand{\joe}[1]{{\protect\color{teal}{\bf [joe: #1]}}}
    \providecommand{\jmh}[1]{{\protect\color{teal}{\bf [joe: #1]}}}
    \providecommand{\david}[1]{{\protect\color{green}{\bf [david: #1]}}}
    \providecommand{\davidmwei}[1]{{\protect\color{pink}{\bf [david wei: #1]}}}
    \providecommand{\kaushik}[1]{{\protect\color{orange}{\bf [kaushik: #1]}}}
    \providecommand{\justin}[1]{{\protect\color{green}{\bf [justin: #1]}}}
    \providecommand{\mingwei}[1]{{\protect\color{rhodamine}{\bf [mingwei: #1]}}}
    \providecommand{\rithvik}[1]{{\protect\color{red}{\bf [rithvik: #1]}}}
    \providecommand{\nc}[1]{{\protect\color{pink}{\bf [nc: #1]}}}
     \providecommand{\accheng}[1]{{\protect\color{olive}{\bf [accheng: #1]}}}
  \else
    \providecommand{\shadaj}[1]{}
    \providecommand{\conor}[1]{}
    \providecommand{\alvin}[1]{}
    \providecommand{\mae}[1]{}
    \providecommand{\joe}[1]{}
    \providecommand{\jmh}[1]{}
    \providecommand{\david}[1]{}
    \providecommand{\davidmwei}[1]{}
    \providecommand{\kaushik}[1]{}
    \providecommand{\justin}[1]{}
    \providecommand{\mingwei}[1]{}
    \providecommand{\rithvik}[1]{}
    \providecommand{\nc}[1]{}
    \providecommand{\accheng}[1]{}
\fi

\ifarxiv
\setcopyright{cc}
\setcctype[4.0]{by}
\else
\fi

\newcommand\vldbdoi{XX.XX/XXX.XX}
\newcommand\vldbpages{XXX-XXX}
\newcommand\vldbvolume{14}
\newcommand\vldbissue{1}
\newcommand\vldbyear{2020}
\newcommand\vldbauthors{\authors}
\newcommand\vldbtitle{\shorttitle} 
\newcommand\vldbavailabilityurl{}
\newcommand\vldbpagestyle{plain} 
\newtheorem{eg}{Example}
\begin{document}
\ifarxiv
\title{Keep CALM and CRDT On}
\fi

\author{Shadaj Laddad}
\affiliation{%
  \institution{University of California, Berkeley}
  \country{}
}
\email{shadaj@cs.berkeley.edu}
\authornote{equal contribution}

\author{Conor Power}
\affiliation{%
  \institution{University of California, Berkeley}
  \country{}
}
\email{conorpower@cs.berkeley.edu}
\authornotemark[1]

\author{Mae Milano}
\affiliation{%
  \institution{University of California, Berkeley}
  \country{}
}
\email{mpmilano@cs.berkeley.edu}

\author{Alvin Cheung}
\affiliation{%
  \institution{University of California, Berkeley}
  \country{}
}
\email{akcheung@cs.berkeley.edu}

\author{Natacha Crooks}
\affiliation{%
  \institution{University of California, Berkeley}
  \country{}
}
\email{ncrooks@cs.berkeley.edu}

\author{Joseph M. Hellerstein}
\affiliation{%
  \institution{University of California, Berkeley}
  \country{}
}
\email{hellerstein@cs.berkeley.edu}

\begin{abstract}
Despite decades of research and practical experience, developers have few tools for programming reliable distributed applications without resorting to expensive coordination techniques. Conflict-free replicated datatypes (CRDTs) are a promising line of work that enable coordination-free replication and offer certain eventual consistency guarantees in a relatively simple object-oriented API. Yet CRDT guarantees extend only to data updates; observations of CRDT state are unconstrained and unsafe. We propose an agenda that embraces the simplicity of CRDTs, but provides richer, more uniform guarantees. We extend CRDTs with a query model that reasons about which queries are safe without coordination by applying monotonicity results from the CALM Theorem, and lay out a larger agenda for developing CRDT data stores that let developers safely and efficiently interact with replicated application state.

\end{abstract}

\maketitle

\ifarxiv

\pagestyle{\vldbpagestyle}

\fi

\section{Introduction}

Consistency is a central theme of distributed computing research, with major implications for 
practitioners. 
Modern cloud-hosted applications are frequently distributed to optimize for latency and availability.
When application state is replicated across the globe, developers often face stark choices 
regarding replica consistency. Strong consistency can be enforced in a general-purpose way at the storage or memory layer via 
classical distributed coordination (consensus, transactions, etc.), but this is often unattractive for latency and availability 
reasons. 
Alternatively, application developers can build on ``weakly'' consistent storage models that do not use coordination; in this case
developers must reason about consistency at the application level. 

The last decade has seen a surge of research interest in reasoning about application consistency, featuring everything from complex formal invariants \cite{quelea} to multi-tiered consistency annotations \cite{mixt, disciplined-inconsistency} to explicit happens-before annotations on operations \cite{c11stuff}. In recent years, one approach has risen above the noise among practitioners: Conflict-Free Replicated Data Types~\cite{CRDTs}. 
CRDTs are provided as an API by a few commercial platforms (e.g., Enterprise Redis,
Akka, Basho Riak \cite{akka,riak,redis}), and have been documented in use by a number of products and services including PayPal, 
League of Legends, and Soundcloud \cite{paypal_crdt,leagueOfLegends,soundcloud}.
There is also are a growing set of open source CRDT packages that have thousands of stars on GitHub \cite{jsonCrdt,automerge,yjs}, 
and blog posts explaining CRDTs to developers in pragmatic, informal terms \cite{bartoszBlog,xiBlog,sephBlog}.

The attractiveness of CRDTs lies in their combination of (1) an easy-to-explain API, and (2) the promise of formal safety guarantees.
Designing a CRDT centers around 
providing a function to \emph{merge} any two replicas, with the requirement that this single function is 
associative, commutative and idempotent (ACI), and defining atomic \emph{operations} that clients can use to update a replica. From the user's perspective, the CRDT's object-oriented API often mimics a familiar collection; many of the CRDTs in the literature are simple adaptations
of well-known data types like Sets and Counters.

The formal safety properties of CRDTs, as originally phrased by Shapiro et al, leverage ``a well defined interface ... [with] mathematically sound rules to guarantee state convergence''~\cite{CRDTs}. This guarantee is achieved via the ACI properties of the merge function. Classic anomalies in eventually consistent systems are caused by reordered, duplicated, or late-arriving updates---none of which can affect the result of an idempotent, commutative, and associative function execution.

But this strong convergence guarantee addresses only state updates and offers no APIs (or guarantees!) 
for \emph{visibility} into the state of a CRDT. Although useful queries are often included in the presentation of CRDT designs, these have no impact on the correctness of the CRDT and are no safer to use than arbitrary queries executed directly on the underlying state. In one of the precursor papers to CRDTs that also proposes ACI merge functions,
Helland and Campbell go as far as noting ironically that READs are ``annoying'' and may not commute with other actions~\cite{buildingOnQuicksand}. 

\vspace{-0.5em}
\begin{eg}[The Potato and the Ferrari, a.k.a.\ Early Read]
    A canonical CRDT is the Two-Phase Set (2P-Set)~\cite{shapiro2011comprehensive}, which is a pair of sets $(A,R)$ that track items 
    to be added ($A$) and removed ($R$). The merge function for two 2P-Sets is defined simply as the 
    pairwise union, $(A_1 \cup A_2, R_1 \cup R_2)$ and is patently ACI. This scheme was used in the well-known Amazon Dynamo 
    shopping cart example~\cite{dynamo}.

    Implicit in this design is a \textbf{query} $Q = A - R$ returning the intended contents 
    of the set. Consider a scenario where a shopper adds a potato and a Ferrari to their cart, then removes the Ferrari, and 
    ``checks out'' by computing the query $Q$. In one or more replicas of the 2P-Set, the checkout 
    request could arrive before the removal of the sports car.  This truly expensive consistency bug arises when the query 
    ``reads'' the state of the 2P-Set ``too early'', before all the removals have eventually arrived. And there is no way to know that
    all the removals have indeed arrived without the coordination that CRDTs supposedly do not need.
\end{eg}

In practice, the soundness of state convergence in CRDTs does not translate to predictable guarantees for computations that examine them. One might say that CRDTs provide \emph{Schrödinger consistency guarantees}: they are guaranteed to be consistent only if they are 
not viewed.

The weak consistency of CRDT queries is not a secret in the research literature~\cite{observable-atomic-consistency,zawirski2015write}, and is mentioned in the initial papers~\cite{CRDTs,shapiro2011comprehensive}. At the same time, bloggers and other developer-facing venues have latched onto the formal language of the 
initial papers (``principled''~\cite{CRDTs}, ``mathematically sound''~\cite{CRDTs}, ``theoretically sound''~\cite{shapiro2011comprehensive}), and sometimes
without caveats. For example, one online article argues that CRDTs ``allow multiple replicas in different regions to mathematically resolve to the same state without coordination ... multiple active copies present accurate views of the shared datasets at low latencies''~\cite{joshiblog}.  This dangerous misread of CRDT guarantees suggests that more work is needed to ensure developers use CRDTs safely.

We believe that the gaps in CRDT guarantees can be addressed on two fronts: (a) defining more precisely what developers must reason about when using CRDTs in their applications and (b) building data systems for CRDTs that automatically manage replication and query execution to deliver stronger consistency guarantees. The unconstrained nature of queries in CRDTs raises an intriguing question: can we develop a more formal query model that makes it possible to precisely define when execution on a single replica yields consistent results?

In this paper, we explore how the CALM (Consistency As Logical Monotonicity) theorem~\cite{calmTheorem}, originally formulated as a definition for consistency in distributed logic programs, can be used as the basis of a query model for CRDTs that delineates queries that can be executed locally from those that require coordination among a quorum. Because monotonicity can be identified as a static property, this view of queries paves the path for a CRDT data system that provides efficient \emph{and} safe execution of queries. Guided by this vision, we map out a research path that weaves together query optimization, storage abstractions, provenance, and more to bring the coordination-free benefits of CRDTs to developers while preserving the consistency guarantees they expect.



\section{An Overview of CRDTs}
\label{sec:crdts}
Most discussions of CRDTs begin by introducing two functionally equivalent representations: \emph{state-based} CRDTs (a.k.a.\ \emph{CvRDTs}) and \emph{op-based} CRDTs (a.k.a.\ \emph{CmRDTs}). Essentially, state-based CRDTs gossip data, while op-based CRDTs gossip logical log records. 

\subsection{State-Based CRDTs}

We begin by reviewing the definition of state-based CRDTs. CvRDTs encapsulate the current $state$ of the replica; let the type of $state$ be called $T$. The API for state-based CRDTs contains three classes of methods, all of which run locally on a single replica's state:

\smallitem{Merge}: \texttt{merge} is a single, required method that takes a value $v$ of type $T$ as input. It 
    combines $state$ with $v$ to generate a value $state'$ of type $T$, and updates 
    itself so that $state = state'$. \emph{Constraint: the \texttt{merge} function must be ACI.}

\smallitem{Operations}: these are methods that clients use to modify $state$. \emph{Constraint: operations must be monotonic with respect to the type $T$.}

\smallitem{Queries}: these are methods that do not modify $state$, but return a result that may be dependent on $state$.
\smallitembot

An example of a 2P-Set CvRDT is shown in Figure~\ref{fig:2pset}. In CvRDTs, nodes gossip their replicas to each other and apply \texttt{merge} upon receipt of gossip. If we focus only on $state$ and \texttt{merge}, a CvRDT is simply a new name for a classical mathematical construct: the \emph{join semi-lattice}. A join semi-lattice is defined in precisely the same way: it is a pair $S = (D, \sqcup)$ where $D$ is a domain (i.e. type) and $\sqcup$ is an operation (called ``join'' or ``least upper bound'') that is ACI.

\begin{figure}
    \begin{lstlisting}
    merge(adds, removes) {
      state.adds = union(state.adds, adds);
      state.removes = union(state.removes, removes);
    } 
    operation add(i) { state.adds = union(state.adds, Set(i)); }
    operation remove(i) { state.removes = union(state.removes, Set(i)); }
    query contents() { return diff(state.adds, state.removes); }
    \end{lstlisting}
    \caption{Pseudocode for a Two-Phase Set CRDT.}
    \label{fig:2pset}
\end{figure}

\begin{figure}
    \centering
    \includegraphics[width=2in]{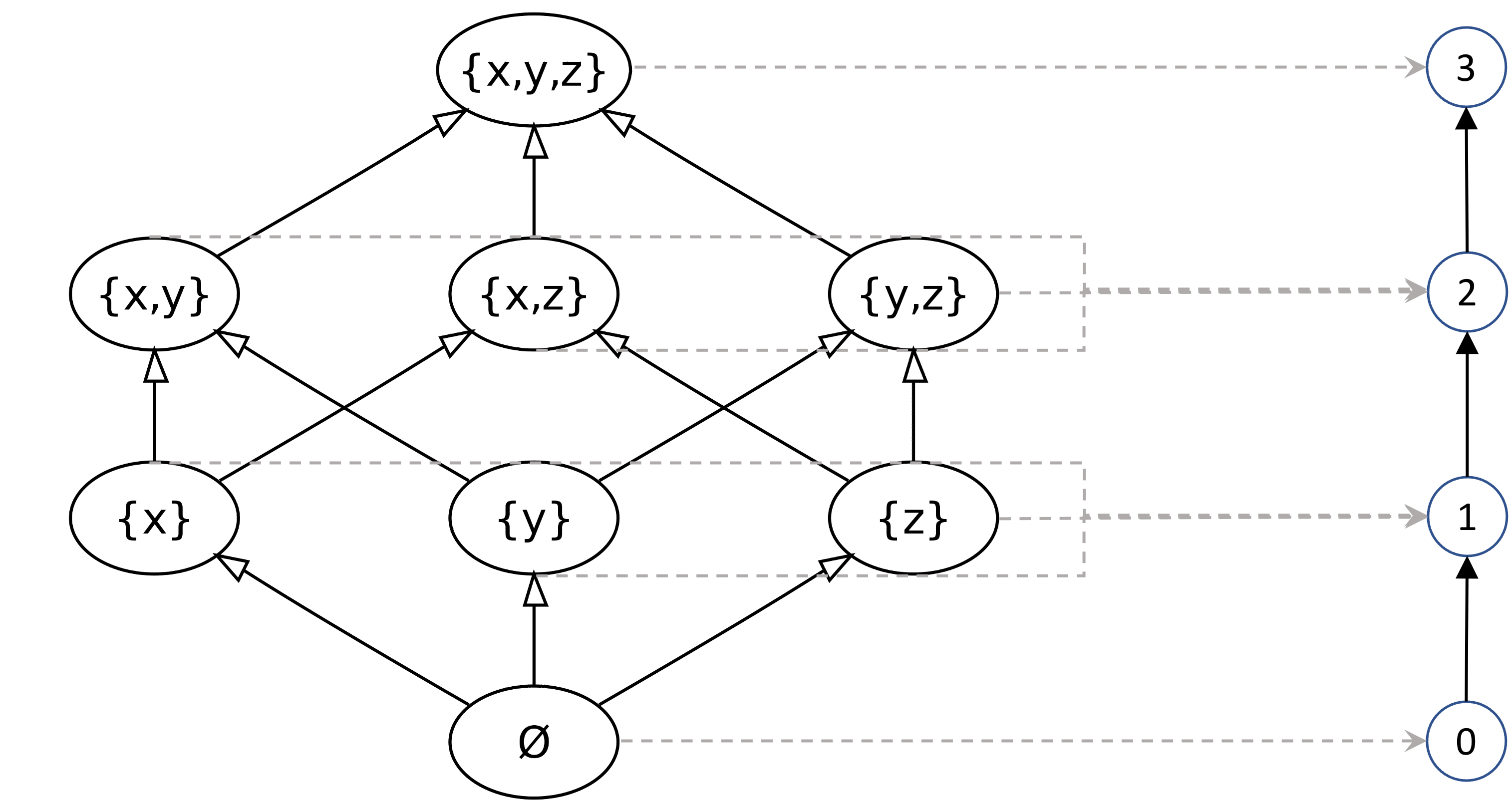}
\caption{Hasse diagrams for G-Set and a cardinality counter, and a monotone function between them (dashed lines).}
\label{fig:hasse}
\end{figure}

A well-known property of a join semi-lattice $S$ is that it is isomorphic to a partial order $\le_S$ on $D$ such that given two elements $s, t \in D$,  $s \le_S t$ iff $s \sqcup t = t$. (We will drop the subscript on $\le$ when it is clear from context.)  The familiar ``Hasse diagrams'' for semi-lattices capture this by laying out the elements of $D$ on a y-axis corresponding to the $\le$ ordering (Figure~\ref{fig:hasse}). Note that this order is partial: two elements $s, t \in D$ may be incomparable: $s \not\le t \wedge t \not\le s$. 

Operations are monotonic with respect to the partial order $\le$: if an operation replaces $state$ with $state'$, we require that $state \le state'$. A good way to enforce this is to forbid operations from modifying $state$ directly, and instead require them to invoke \texttt{merge} to perform the state update---the ACI properties of \texttt{merge} then ensure monotonic updates. Viewed through that lens, CvRDTs only update their state through the ACI \texttt{merge} function, and are precisely join semi-lattices~\cite{katara}.

Note that although queries are included in the API surface of CvRDTs, they are unconstrained and may perform arbitrary computations on the underlying state. Therefore, the choice of queries included with a CRDT design have no effect on the \emph{safety} of observations through them---their consistency guarantees are no stronger than a single query that just emits the internal state of the CRDT.


\subsection{Op-Based CRDTs and Compression}
The idea of op-based CRDTs (CmRDTs) is to gossip logs of operations rather than state. 
Each given replica $r$ applies its operations sequentially, so upon gossip we require that another replica $r'$ applies the log records from $r$ in an order that produces the same final state as $r$. In typical deployments of op-based CRDTs, this restriction is imposed by requiring the network between replicas to guarantee causal delivery. As a result, the logs accumulated at every replica follow the standard \emph{happens-before} partial order defined by Lamport~\cite{lamportTimestamps}.

But as noted above, a partial order like happens-before is isomorphic to a join semi-lattice! That means that we can capture the CmRDT log as a CvRDT. One simple way to do this is to hold the DAG corresponding to the partial order: this is simply a ``grow-only'' set---whose only operation is \texttt{add}---which holds edges of a DAG. Each directed edge connects two operations, and indicates that the operations happened in the order of the edge. (Another solution would be to use CvRDTs for vector clocks.) A natural query over such a CvRDT is to ``play the log'': i.e. produce a topological sort of the log and apply the operations in the resulting order.

CmRDTs can therefore be considered a gossip compression technique that maps instances of one CvRDT (say a 2P-Set as in Figure~\ref{fig:2pset}) to another CvRDT (a partially ordered log of \texttt{add} and \texttt{remove} operations). 
Hence CmRDTs are arguably a specific form of a CvRDT for partially-ordered logs, with various tricks applied to suit the nature of the operations being logged. In the remainder of our discussion, we focus on the clean lattice-based CvRDT API, and use the CRDT acronym to mean CvRDTs.

\subsection{A Snapshot of CRDTs in Practice}
This model of CRDTs has gained traction in the industry across a wide range of applications ranging from high-scale backend logic to client-side collaborative state. Before we dive into our vision for extensions to CRDTs that make them safer to use, let us briefly explore the ways they are already being applied.

CRDT designs in use today largely fall into two buckets: the core CRDTs from the early literature designed to mimic classic data structures~\cite{shapiro2011comprehensive} and more advanced CRDTs focused on replicating documents for collaborative editing~\cite{logoot, wooki}. Across a variety of languages, developers have created several libraries~\cite{rust_crdt,automerge} that provide high-quality implementations of the core CRDTs. These CRDTs have also become adopted as building blocks that can be used by distributed systems developers through systems like Akka~\cite{akka}, Dynamo~\cite{dynamo}, and Redis~\cite{redis}, which all provide CRDTs as built-in data structures. There are also well documented examples of industry players building systems on top of CRDTs, such as PayPal~\cite{paypal_crdt} and League of Legends~\cite{leagueOfLegends}. We even see examples of CRDTs used inside of databases such as in FlightTracker~\cite{flighttracker} at Facebook, which uses CRDTs to provide stronger consistency guarantees in the TAO data store.

A significant portion of recent CRDT research has focused on \emph{collaborative editors}, which have latency and fault-tolerance challenges that CRDTs are well-suited to address. The research in this space is primarily interested in representing the variety of ways that users interact with text documents, such as text insertion, cut/paste to different locations, and formatting layers. A lot of recent creativity in the CRDT space has gone into this domain, resulting in designs such as Peritext~\cite{peritext}. 

\section{Toward a Query Model for CRDTs}
\label{section:query_model}

In Section~1 we argued that CRDTs have gained interest for their combination of 
safety, efficiency, and simplicity. In that spirit, our desiderata for a good CRDT query model are:

    \smallitem{Safety:} Queries should be sequentially consistent, regardless of the replica at which they are evaluated.

    \smallitem{Efficiency:} Queries should be evaluated locally without coordination whenever possible.
    
    \smallitem{Simplicity:} The query model should be easy for developers to reason about.
    
    \smallitembot


\subsection{Example Queries}
Let's look at some examples of queries that can and cannot be executed without coordination while satisfying sequential consistency.

\begin{eg}[A Boolean Threshold Query Over a Grow-Only Set]
    One of the simplest possible CRDTs is the Grow-Only Set (G-Set)~\cite{shapiro2011comprehensive}. It is a set $S$ with an operation that can add elements to $S$ and a merge function that is the set union, $S_1 \cup S_2$. 

    Say we want to determine whether the number of gift card purchases that are over \$100 in a set of transactions has exceeded 50 items (similar to the threshold functions in LVars~\cite{lvars}): 
    \begin{lstlisting}
    query suspicious_activity() { 
      if (cardinality([
        txn for txn in state
        if txn.type == "GIFTCARD" and txn.amount > 100
      ]) > 50):
        return true else ABORT;
    }
    \end{lstlisting}
    \label{eg:thresh}
\end{eg}
    Note that the \texttt{suspicious\_activity} query returns either \texttt{true} or aborts (signifying ``unknown''). Perhaps surprisingly, executing this query on a local replica will always produce a sequentially consistent result, even without coordination!  This is because each replica of a CRDT effectively represents an {\it under-approximation} of some true global state; that is, each individual replica has seen some {\it subset} of the updates which have entered the system at any given time.  Thus, the true ``global'' state of the system can be derived from any individual replica's local state by adding in some number of missing update operations.  This query's {\tt true} result {\it cannot be changed} by any subsequent updates; thus, for the purposes of {\it this query}, our replica's local state reflects the same information as a true ``global'' state would contain, and can serialize after that global state. If the query aborts, we learn nothing; aborted queries are not considered when determining sequential consistency.
    
    
    Let's return to our Potato/Ferrari example from the introduction. In this example, the query we wanted to ask, $Q = A - R$, would not yield a consistent result when executed locally. We again are guaranteed that \emph{eventually} the local state will converge to the global state and give us the correct answer to the query, by the eventually consistent properties of CRDT operations. However, \emph{unless we can ensure that our local state is equal to the global state via coordination}, we cannot determine that the result of our local query matches the result of the global query. We could be missing the effect of an operation locally (e.g., \texttt{remove(`Ferrari')}) and we would output an incorrect result.
    
    Consider a third query, again over the 2P-Set. This time we want to rate-limit a shopping cart user by detecting whether the number of actions they have taken exceeds 100. Our query is $Q = |A| + |R| > 100$. This query can computed locally without coordination by the same reasoning as \texttt{suspicious\_activity}: since $A$ and $R$ individually are guaranteed to increase in size, we know that the $global(|A| + |R|)\geq local(|A| + |R|)$.

\subsection{What's Going on Here? ... Monotonicity!}
\label{section:monotone}

What is going on here? Some queries are consistent without coordination, but others require a global view of the system to be correct? The difference between these queries is that the threshold queries are \textbf{monotone} queries with respect to the CRDT state and its partial order, whereas the shopping cart checkout query is \textbf{non-monotone}. This distinction may be familiar to the reader from the CALM Theorem~\cite{calmTheorem,keepingCalm}, which proved that programs are convergent without coordination iff they are monotone. 

The CALM Theorem is framed in terms of monotonicity over logic on relations---but it applies equally well to the CRDT domain! Per Section~\ref{sec:crdts}, we know that both the merge and update methods of a CRDT monotonically increase the value of the CRDT's state with respect to its partial order. 
We can define a monotone query as any whose output is monotone with respect to ordering of the CRDT. That is, given a join semi-lattice $S = (D, \sqcup)$ as the state of the CRDT, a query $Q$ is monotone if $\forall i, j \in D: i \le j, \; Q(i) \implies Q(j)$. By the CALM Theorem, monotone queries over CRDTs are exactly the queries that only need a local view of the system to be correct!



Monotone queries meet all the criteria we outlined for a good query model. They are able to achieve safety and efficiency for queries over CRDTs. The CALM Theorem tells us that not only do they meet these criteria (if), but they are the only queries that meet this criteria (only if). Further, monotone queries offer composition through monotone functions. As previously observed (Bloom$^L$~\cite{bloomL}, Lasp~\cite{lasp}, Datafun~\cite{datafun}) this compositionality allows construction of complex systems out of CRDT primitives, and is highly amenable to programming language tools. The space of monotone queries is quite large; for example, four of the five operators of relational algebra are monotone: selection, projection, union, and intersection. Only set difference is non-monotone. Observe that a pipeline composing monotone functions will always give a monotone function end-to-end, but if the pipeline contains any non-monotone function then the end-to-end-computation will be non-monotone.

Perhaps the most important property, since it focuses on the adoption barriers for CRDTs, is the simplicity of our query model. We believe that query monotonicity should be understandable to anyone who understands basic CRDTs. The state in standard CRDT examples is either sets or counters, both of which have simple, intuitive definitions of monotonicity. Moreover, the definition of CRDTs requires developers to understand monotonicity with respect to state updates, so it should be reasonable for developers to extend this reasoning to include queries as well. Because monotonicity can be syntactically identified in many existing query languages, including SQL, we are optimistic that developer tools can help guide the creation of monotonic queries.

\subsection{Beyond Monotonicity}
\label{section:non_monotone}

Monotonic queries are the natural query model for CRDTs; their resilience to update re-ordering mirrors the convergent nature of updates within a CRDT. We acknowledge, however, that there are large classes of queries performed on CRDTs which are {\it not} monotonic. A simple example of a non-monotonic query is set difference ($Q = A - R$), seen in our shopping cart. In non-monotonic queries, missing global information can make results appear to go {\it backwards} over time, making it impossible to safely make decisions based on the results of these queries.

So what are developers to do when they need one of these non-monotone queries? The simple and safe solution is to coordinate! Queries executed against global state, after all, will always be correct. The choice of coordination falls into a classic spectrum for distributed databases outlined by Bernstein and Goodman~\cite{bernstein1981concurrency}: write-one read-all, write-majority read-majority, or write-all read-one. Each of these strategies, however, is still improved by the use of CRDTs: with CRDTs, only non-monotone queries need to be ordered, with respect to {\it sets} of updates. As update operations commute, there is no need to coordinate in order to sequence update operations with each other. 


Developers building on CRDTs retain the option to perform a local, but stale query on a single replica. The resulting application-level considerations align well with those established by high-scale systems developed in industry, such as Google's Zanzibar authorization system~\cite{zanzibar}, which offers APIs for retrieving safe, up-to-date results or recent, but potentially stale ones. For applications that can tolerate out-of-date results, with a staleness distribution determined by the gossip protocol, local non-monotone queries offer a fast path that can reduce the overall latency perceived by the user.

\section{Enabling Fast and Safe CRDT Systems}
\label{section:puttingItTogether}

Equipped with our distinction of monotone queries, we believe that developers will be able to apply CRDTs in new ways by developing complex applications on top of replicated state. By reducing query correctness to a simple property, monotonicity, developers can reason about the correctness of their data architectures and know the pitfalls to avoid when creating CRDT-backed systems. As more developers write software that depends on CRDTs in complex ways, it is critical that the research community explores methods that help developers utilize CDRTs robustly and efficiently. To this end, we propose a shift in perspective from an object-oriented view of CRDTs to a “database” view of them: breaking CRDTs up into a query model and a data store that separates their logical and physical representations. This shift in perspective brings us to many impactful research problems in data management for CRDTs. 

\subsection{A Query Language For CRDTs}

In Section~\ref{section:query_model}, we outlined a formal query model for CRDTs that uses monotonicity as a key property to determine how the query should be executed to guarantee consistent results. Our next challenge is to identify how this model can be mapped to a practical language that developers can use to declare the observations they want to make on CRDTs.

What we need in a query language is a set of rich expressions that can manipulate the lattice structures used inside CRDTs, and a syntax that makes it easy for developers (and computers) to identify when a query is monotone and can be efficiently executed. One promising choice is to develop a dialect of the query language most developers are already familiar with: SQL! Recent theoretical work~\cite{khamisPODS22} demonstrated an extension of relational queries to operate over semi-rings. We are currently working on a similar formalization for extending relational algebra to semi-lattices.

There are two major benefits of using the SQL language for queries over CRDTs. First, it is clear syntactically in SQL queries whether the query is monotone or not, which will help developers design efficient programs and reason about them. Second, we already know how to build optimizers for SQL that take advantage of monotonicity! Streaming joins and barrier stages are examples of an optimizer leveraging its knowledge of monotonicity and non-monotonicity respectively in the dataflow graph.

Using relational-style languages to query CRDTs also fits well into recent research exploring alternate models of CRDTs. In particular, there has been a recent push to define CRDTs as Datalog queries over sets of operations being gossiped across nodes~\cite{kleppmannDatalog}. In such a model, issuing queries in a relational-style language naturally fits into the execution model, and opens up the opportunity for further end-to-end optimizations such as using incremental view maintenance to identify efficient ways to propagate the effects of operations to queries. 



By defining a query language that allows developers to safely view the state of a CRDT, we arrive at an interaction model that is distinct from the object-oriented, in-memory CRDTs used today. Our model of CRDTs includes application-specific operations, which are fundamental to proving convergence of the CRDT, but the lack of predefined queries deviates significantly from the classic object-oriented model. Our CRDTs can be viewed as effectively having one open-ended endpoint that all queries go through. 

\subsection{A Data Management System for CRDTs}
Separating the capabilities of CRDTs from the object-oriented interface they are typically wrapped in raises an intriguing opportunity: operations and our query language can become an interface between application logic and an independent CRDT data store that manages all aspects of the CRDT life-cycle. We believe that this approach can both increase the ease of use of CRDTs, by shifting the responsibility of reasoning about consistency to the store, and improve the efficiency of applications built on CRDTs, since data stores can make advanced optimization decisions based on the dynamic workload. Compared to existing data stores that support CRDTs but do not provide query APIs~\cite{redis, tardis}, our monotonic query model enables rich observations of CRDTs with strong consistency guarantees.

In our vision, a CRDT data store sits in the application stack at a similar level as a traditional relation database or key-value store. Application services can interact with the CRDT store over the network, using a protocol that can be easily implemented by several languages so that heterogeneous application implementations can interact on shared CRDT state. By deploying the CRDT store as a separate networked service, our approach also enables serverless applications to build on top of replicated state~\cite{cloudburst}, since the replicas will be maintained separately from the ephemeral execution nodes.

Shifting CRDTs from being baked into application logic to being managed by an external service does raise a challenge in the extensibility with respect to the available data types. While today's CRDTs are defined as regular data types in the application language and are immediately usable, bringing custom types to a CRDT data store requires a pluggable interface to provide the data store with an implementation of the CRDT. We believe that this can be achieved with a lightweight extension API, similar to Postgres extensions~\cite{postgres}, that uses a foreign-function interface to inject custom CRDTs. Because we only rely on the core CRDT properties of monotonic state and convergent operations, the system can immediately make the CRDT available as long as the developer certifies it satisfies these properties.

\subsection{Optimization Opportunities in CRDT Stores}
The abstraction of placing CRDTs in a store separate from the application reveals a range of opportunities for research that optimizes how the CRDTs are stored and queried.
Having applied our first major database trick, a query model, the next natural one to apply is the separation of logical and physical data representations. By separating these for the CRDT data model, we open up many research directions for optimizing the physical layer of CRDTs in the DBMS. An increasingly popular research topic for CRDTs is how to minimize their memory footprint~\cite{dson}. Recent work applied columnar compression techniques from databases to collaborative editor CRDTs~\cite{kleppmanColumnarCRDT}. The more general question of how to automatically find compressed representations of CRDTs is open.

The other approach to minimizing the memory footprint that is ripe for automated management is garbage collection: when can you delete or compact older operations in the CRDT? In today's applications, where CRDTs are freely passed around as regular objects, it can be difficult to trace down all replicas. But when all CRDT state is managed by the data store, garbage collection can be safely employed because it is clear where replicas of each CRDT lie. Garbage collection requires advanced program analyses to determine when components of CRDT state become inaccessible. We are especially excited about the opportunity to apply program synthesis and automated verification research to discover strategies for garbage collection.

Finally, there are several enticing research topics focused on how the effects of operations are propagated between replicas. The classic algorithm for gossip with state-based CRDTs calls for the \emph{entire} state of each replica to be sent over the network. For large CRDT instances, this places a significant networking burden that can increase staleness. But with bookkeeping that captures which versions of the state have already been gossiped, we can instead propagate smaller deltas that capture the effects of new operations. There are several intriguing research directions in this space, such as identifying compact data structures for the bookkeeping, the selection of what delta information to gossip, how to batch those deltas, and what gossip architecture and frequency to use. 

\subsection{Tradeoffs for Non-Monotone Queries}
Although monotone queries make the most of CRDTs by entirely eliminating coordination, not all business logic can be expressed strictly in terms of such queries. Consistently executing non-monotone queries requires the CRDT data store to introduce coordination among replicas. However, not all hope for low-latency queries is lost!

As discussed in Section~\ref{section:non_monotone}, datastores would only need to coordinate in anticipation of a non-monotone query.  With accurate metrics and a model to predict both the frequency of these queries and the data such queries will touch, future datastores can perform much of the work of reaching consistency in advance of this non-monotone query, shifting into and out of a synchronize-on-update model accordingly (as previously observed in \cite{redblue-automating, redblue}). We further note, however, that the {\it metrics themselves} are likely to be monotone---and thus the work of determining when to coordinate is itself amenable to our monotone query model, and can therefore avoid synchronization. 


\subsubsection{Weakening consistency for non-monotonicity}

Some applications may prefer to avoid coordination entirely, despite the fact that coordination in the presence of non-monotonic queries is essential for maintaining consistency.  These applications have essentially decided that {\it weak consistency} is tolerable for their purposes.  Much previous work exists in the space of weakly-consistent datastores, both in terms of how to best supply usable weak consistency to the user \cite{tardis,cure,cops,cassandra,anna} and how to help the user decide {\it when} weakening consistency may be appropriate \cite{gallifrey, quelea,strongEnough}.

Our framework of monotonic queries over CRDTs can enhance these existing approaches.  Most obviously, it can recognize patterns of weakly-consistent reads which form monotonic queries, and can thus allow such patterns without sacrificing consistency.  More subtly, building CRDTs into a database enables the use of lineage techniques that have been well-studied in existing database systems. In particular, a CRDT-and-monotonicity-aware dataflow technique would be capable of determining the potential {\it downstream effects} of a weakly-consistent query, providing database users and administrators alike with valuable insights into the consistency of not just their data, but the observations {\it derived} from that data.

This information has particular use when applied to the {\it apologies} strategy first proposed by Helland and Campbell in \cite{buildingOnQuicksand}. With apologies, potentially-inconsistent observations are accompanied by {\it compensating actions}, which are intended to clean up any negative effects of weak consistency.  By leveraging lineage tracing, a CRDT-enabled database could automatically determine when such apologies are necessary, prompting the application accordingly. Indeed, this strategy is already present in PL-focused solutions such as \cite{gallifrey}.

\section{Related Work}

CRDTs are well-studied in the research community; seminal citations mentioned earlier include~\cite{CRDTs, shapiro2011comprehensive}, as well as many papers on collaborative-text CRDTs~\cite{wooki,logoot,peritext,xiBlog}.
We choose to focus on CRDTs due to their rising popularity; this framing drives our choice to explore queries over CRDTs, and in turn excludes lines of work which focus on safely directly observing weakly-consistent shared memory such as \cite{c11stuff,quelea,pileus}.  In particular, solutions which rely on causal \cite{cops}, linearizable \cite{linearizability}, or other explicit consistency models stronger than eventual consistency (such as \cite{mixt,disciplined-inconsistency,tardis,ralinearizability,epsilonSerializability,mdcc,redblue,observable-atomic-consistency}), were set aside in our discussion.

These papers attempt to reduce the number of anomalies that may emerge from weakly-consistent applications by eliminating certain reorderings on some operations; in contrast, we seek a simple unifying principle by which a developer can consider a CRDT observation to be reliable, {\it without} reasoning about varying classes of inconsistency under which it may be read.  Some work, such as \cite{redblue-automating, mdcc}, uses CRDT-like reasoning to automate the choice of consistency model; we believe that these papers fit with our goal of discovering classes of monotonic observations safe to perform with weak consistency, and we seek to extend them here. 

While we focus on how developers \emph{use} CRDTs, there is a wide range of research on how these data types are \emph{designed}. Recent work has explored how CRDT designs can be generated using program synthesis techniques~\cite{katara}, which significantly reduces the burden of designing new types for custom application logic. In addition, a few replicated data types have been proposed as alternatives to CRDTs, such as ECROs~\cite{ecros} and MRDTs~\cite{mrdts}, with the focus of making it easier for developers to express the semantics of the merge logic. Like traditional CRDTs, however, these are also focused on convergent updates to the data type, and leave queries unconstrained. Other work, such as \cite{observable-atomic-consistency}, introduce total order to certain CRDT operations in order to allow consistent observations; in contrast, we attempt to identify a class of {\it monotonic} observations which do not require total order. Similarly, other work solves the consistent read problem in ways reminiscent of escrow transactions~\cite{escrow} or single-master replication~\cite{homeostasis,warranties}.

Perhaps the most related work comes from the programming languages and databases space, with languages such as Gallifrey, LVars, Lasp, Datafun, and Bloom$^L$ all providing capabilities for constraining monotonic logic \cite{bloomL,gallifrey,lvars,datafun,lasp}. In Section 4, we discuss a SQL-like query language with a database optimizer and intelligent storage layer. Almost all of the research directions we outline in that setting also apply to the compiler, runtime, and storage layer for monotonicity-aware DSLs like these.

We believe that lighter-weight solutions than rewriting applications in a new DSL or query language will also play an important role in safe adoption of CRDTs. One promising such solution is type annotations in existing languages along the lines of Blazes~\cite{blazes}. In a DSL designed for CRDT queries, similar type systems could be used to enforce monotonicity while supporting complex query logic. Other angles from the programming languages and software engineering communities include formal verification and fuzzing, which can assist developers designing CRDTs by automatically checking the core correctness properties.



\section{Conclusion}
CRDTs are simple, and beginning to see adoption among developers—an important signal for database researchers. 
CRDTs on their own lack power. However, the research literature abounds with results---such as the CALM theorem---which when combined with CRDTs open new frontiers to researchers and developers alike. The next generation of challenges in this space will arise at the seams between foundational research and practical concerns. These challenges will require research spanning data management, distributed systems, query models, and programming languages.

\ifarxiv
\clearpage
\else
\fi

\bibliographystyle{ACM-Reference-Format}
\bibliography{bibliography,akcheung}


\begin{thebibliography}{64}


\ifx \showCODEN    \undefined \def \showCODEN     #1{\unskip}     \fi
\ifx \showDOI      \undefined \def \showDOI       #1{#1}\fi
\ifx \showISBNx    \undefined \def \showISBNx     #1{\unskip}     \fi
\ifx \showISBNxiii \undefined \def \showISBNxiii  #1{\unskip}     \fi
\ifx \showISSN     \undefined \def \showISSN      #1{\unskip}     \fi
\ifx \showLCCN     \undefined \def \showLCCN      #1{\unskip}     \fi
\ifx \shownote     \undefined \def \shownote      #1{#1}          \fi
\ifx \showarticletitle \undefined \def \showarticletitle #1{#1}   \fi
\ifx \showURL      \undefined \def \showURL       {\relax}        \fi
\providecommand\bibfield[2]{#2}
\providecommand\bibinfo[2]{#2}
\providecommand\natexlab[1]{#1}
\providecommand\showeprint[2][]{arXiv:#2}

\bibitem[\protect\citeauthoryear{Akkoorath, Tomsic, Bravo, Li, Crain, Bieniusa,
  Pregui{\c{c}}a, and Shapiro}{Akkoorath et~al\mbox{.}}{2016}]%
        {cure}
\bibfield{author}{\bibinfo{person}{Deepthi~Devaki Akkoorath},
  \bibinfo{person}{Alejandro~Z Tomsic}, \bibinfo{person}{Manuel Bravo},
  \bibinfo{person}{Zhongmiao Li}, \bibinfo{person}{Tyler Crain},
  \bibinfo{person}{Annette Bieniusa}, \bibinfo{person}{Nuno Pregui{\c{c}}a},
  {and} \bibinfo{person}{Marc Shapiro}.} \bibinfo{year}{2016}\natexlab{}.
\newblock \showarticletitle{Cure: Strong semantics meets high availability and
  low latency}. In \bibinfo{booktitle}{\emph{2016 IEEE 36th International
  Conference on Distributed Computing Systems (ICDCS)}}. IEEE,
  \bibinfo{pages}{405--414}.
\newblock


\bibitem[\protect\citeauthoryear{Alvaro, Conway, Hellerstein, and Maier}{Alvaro
  et~al\mbox{.}}{2014}]%
        {blazes}
\bibfield{author}{\bibinfo{person}{Peter Alvaro}, \bibinfo{person}{Neil
  Conway}, \bibinfo{person}{Joseph~M Hellerstein}, {and} \bibinfo{person}{David
  Maier}.} \bibinfo{year}{2014}\natexlab{}.
\newblock \showarticletitle{Blazes: Coordination analysis for distributed
  programs}. In \bibinfo{booktitle}{\emph{2014 IEEE 30th International
  Conference on Data Engineering}}. \bibinfo{pages}{52--63}.
\newblock


\bibitem[\protect\citeauthoryear{Ameloot, Neven, and Van~den Bussche}{Ameloot
  et~al\mbox{.}}{2013}]%
        {calmTheorem}
\bibfield{author}{\bibinfo{person}{Tom~J Ameloot}, \bibinfo{person}{Frank
  Neven}, {and} \bibinfo{person}{Jan Van~den Bussche}.}
  \bibinfo{year}{2013}\natexlab{}.
\newblock \showarticletitle{Relational transducers for declarative networking}.
\newblock \bibinfo{journal}{\emph{Journal of the ACM (JACM)}}
  (\bibinfo{year}{2013}), \bibinfo{pages}{1--38}.
\newblock


\bibitem[\protect\citeauthoryear{Arntzenius and Krishnaswami}{Arntzenius and
  Krishnaswami}{2016}]%
        {datafun}
\bibfield{author}{\bibinfo{person}{Michael Arntzenius} {and}
  \bibinfo{person}{Neelakantan~R Krishnaswami}.}
  \bibinfo{year}{2016}\natexlab{}.
\newblock \showarticletitle{Datafun: a functional Datalog}. In
  \bibinfo{booktitle}{\emph{Proceedings of the 21st ACM SIGPLAN International
  Conference on Functional Programming}}. \bibinfo{pages}{214--227}.
\newblock


\bibitem[\protect\citeauthoryear{Bernstein and Goodman}{Bernstein and
  Goodman}{1981}]%
        {bernstein1981concurrency}
\bibfield{author}{\bibinfo{person}{Philip~A Bernstein} {and}
  \bibinfo{person}{Nathan Goodman}.} \bibinfo{year}{1981}\natexlab{}.
\newblock \showarticletitle{Concurrency control in distributed database
  systems}.
\newblock \bibinfo{journal}{\emph{ACM Computing Surveys (CSUR)}}
  (\bibinfo{year}{1981}), \bibinfo{pages}{185--221}.
\newblock


\bibitem[\protect\citeauthoryear{Bourgon}{Bourgon}{2014}]%
        {soundcloud}
\bibfield{author}{\bibinfo{person}{Peter Bourgon}.}
  \bibinfo{year}{2014}\natexlab{}.
\newblock \showarticletitle{Roshi: a CRDT system for timestamped events}.
\newblock  (\bibinfo{date}{May} \bibinfo{year}{2014}).
\newblock
\urldef\tempurl%
\url{https://developers.soundcloud.com/blog/roshi-a-crdt-system-for-timestamped-events}
\showURL{%
\tempurl}


\bibitem[\protect\citeauthoryear{Community}{Community}{2022}]%
        {redis}
\bibfield{author}{\bibinfo{person}{The~Redis Community}.}
  \bibinfo{year}{2022}\natexlab{}.
\newblock \bibinfo{title}{Redis}.
\newblock
\newblock
\urldef\tempurl%
\url{https://redis.io}
\showURL{%
\tempurl}


\bibitem[\protect\citeauthoryear{Conway, Marczak, Alvaro, Hellerstein, and
  Maier}{Conway et~al\mbox{.}}{2012}]%
        {bloomL}
\bibfield{author}{\bibinfo{person}{Neil Conway}, \bibinfo{person}{William~R.
  Marczak}, \bibinfo{person}{Peter Alvaro}, \bibinfo{person}{Joseph~M.
  Hellerstein}, {and} \bibinfo{person}{David Maier}.}
  \bibinfo{year}{2012}\natexlab{}.
\newblock \showarticletitle{Logic and Lattices for Distributed Programming}. In
  \bibinfo{booktitle}{\emph{Proceedings of the Third ACM Symposium on Cloud
  Computing}}.
\newblock


\bibitem[\protect\citeauthoryear{Crooks, Pu, Estrada, Gupta, Alvisi, and
  Clement}{Crooks et~al\mbox{.}}{2016}]%
        {tardis}
\bibfield{author}{\bibinfo{person}{Natacha Crooks}, \bibinfo{person}{Youer Pu},
  \bibinfo{person}{Nancy Estrada}, \bibinfo{person}{Trinabh Gupta},
  \bibinfo{person}{Lorenzo Alvisi}, {and} \bibinfo{person}{Allen Clement}.}
  \bibinfo{year}{2016}\natexlab{}.
\newblock \showarticletitle{TARDiS: A Branch-and-Merge Approach To Weak
  Consistency}. In \bibinfo{booktitle}{\emph{Proceedings of the 2016
  International Conference on Management of Data}}.
  \bibinfo{pages}{1615–1628}.
\newblock


\bibitem[\protect\citeauthoryear{De~Porre, Ferreira, Pregui\c{c}a, and
  Gonzalez~Boix}{De~Porre et~al\mbox{.}}{2021}]%
        {ecros}
\bibfield{author}{\bibinfo{person}{Kevin De~Porre}, \bibinfo{person}{Carla
  Ferreira}, \bibinfo{person}{Nuno Pregui\c{c}a}, {and} \bibinfo{person}{Elisa
  Gonzalez~Boix}.} \bibinfo{year}{2021}\natexlab{}.
\newblock \showarticletitle{ECROs: Building Global Scale Systems from
  Sequential Code}.
\newblock \bibinfo{journal}{\emph{Proc. ACM Program. Lang.}}
  (\bibinfo{year}{2021}).
\newblock


\bibitem[\protect\citeauthoryear{DeCandia, Hastorun, Jampani, Kakulapati,
  Lakshman, Pilchin, Sivasubramanian, Vosshall, and Vogels}{DeCandia
  et~al\mbox{.}}{2007}]%
        {dynamo}
\bibfield{author}{\bibinfo{person}{Giuseppe DeCandia}, \bibinfo{person}{Deniz
  Hastorun}, \bibinfo{person}{Madan Jampani}, \bibinfo{person}{Gunavardhan
  Kakulapati}, \bibinfo{person}{Avinash Lakshman}, \bibinfo{person}{Alex
  Pilchin}, \bibinfo{person}{Swaminathan Sivasubramanian},
  \bibinfo{person}{Peter Vosshall}, {and} \bibinfo{person}{Werner Vogels}.}
  \bibinfo{year}{2007}\natexlab{}.
\newblock \showarticletitle{Dynamo: Amazon's highly available key-value store}.
\newblock \bibinfo{journal}{\emph{ACM SIGOPS operating systems review}}
  (\bibinfo{year}{2007}), \bibinfo{pages}{205--220}.
\newblock


\bibitem[\protect\citeauthoryear{Doherty, Dongol, Wehrheim, and
  Derrick}{Doherty et~al\mbox{.}}{2019}]%
        {c11stuff}
\bibfield{author}{\bibinfo{person}{Simon Doherty}, \bibinfo{person}{Brijesh
  Dongol}, \bibinfo{person}{Heike Wehrheim}, {and} \bibinfo{person}{John
  Derrick}.} \bibinfo{year}{2019}\natexlab{}.
\newblock \showarticletitle{Verifying C11 programs operationally}. In
  \bibinfo{booktitle}{\emph{Proceedings of the 24th Symposium on Principles and
  Practice of Parallel Programming}}. \bibinfo{pages}{355--365}.
\newblock


\bibitem[\protect\citeauthoryear{Gentle}{Gentle}{2022}]%
        {sephBlog}
\bibfield{author}{\bibinfo{person}{Seph Gentle}.}
  \bibinfo{year}{2022}\natexlab{}.
\newblock \bibinfo{title}{5000x faster CRDTs: An Adventure in Optimization}.
\newblock
\newblock
\newblock
\shownote{\url{https://josephg.com/blog/crdts-go-brrr//}, retrieved June 1,
  2022.}


\bibitem[\protect\citeauthoryear{Gotsman, Yang, Ferreira, Najafzadeh, and
  Shapiro}{Gotsman et~al\mbox{.}}{2016}]%
        {strongEnough}
\bibfield{author}{\bibinfo{person}{Alexey Gotsman}, \bibinfo{person}{Hongseok
  Yang}, \bibinfo{person}{Carla Ferreira}, \bibinfo{person}{Mahsa Najafzadeh},
  {and} \bibinfo{person}{Marc Shapiro}.} \bibinfo{year}{2016}\natexlab{}.
\newblock \showarticletitle{'Cause I'm strong enough: Reasoning about
  consistency choices in distributed systems}. In
  \bibinfo{booktitle}{\emph{Proceedings of the 43rd Annual ACM SIGPLAN-SIGACT
  Symposium on Principles of Programming Languages}}.
  \bibinfo{pages}{371--384}.
\newblock


\bibitem[\protect\citeauthoryear{Helland and Campbell}{Helland and
  Campbell}{2009}]%
        {buildingOnQuicksand}
\bibfield{author}{\bibinfo{person}{Pat Helland} {and} \bibinfo{person}{David
  Campbell}.} \bibinfo{year}{2009}\natexlab{}.
\newblock \showarticletitle{Building on quicksand}.
\newblock \bibinfo{journal}{\emph{arXiv preprint arXiv:0909.1788}}
  (\bibinfo{year}{2009}).
\newblock


\bibitem[\protect\citeauthoryear{Hellerstein and Alvaro}{Hellerstein and
  Alvaro}{2020}]%
        {keepingCalm}
\bibfield{author}{\bibinfo{person}{Joseph~M Hellerstein} {and}
  \bibinfo{person}{Peter Alvaro}.} \bibinfo{year}{2020}\natexlab{}.
\newblock \showarticletitle{Keeping CALM: when distributed consistency is
  easy}.
\newblock \bibinfo{journal}{\emph{Commun. ACM}} (\bibinfo{year}{2020}),
  \bibinfo{pages}{72--81}.
\newblock


\bibitem[\protect\citeauthoryear{Herlihy and Wing}{Herlihy and Wing}{1990}]%
        {linearizability}
\bibfield{author}{\bibinfo{person}{Maurice~P Herlihy} {and}
  \bibinfo{person}{Jeannette~M Wing}.} \bibinfo{year}{1990}\natexlab{}.
\newblock \showarticletitle{Linearizability: A correctness condition for
  concurrent objects}.
\newblock \bibinfo{journal}{\emph{ACM Transactions on Programming Languages and
  Systems (TOPLAS)}} (\bibinfo{year}{1990}), \bibinfo{pages}{463--492}.
\newblock


\bibitem[\protect\citeauthoryear{Hoff}{Hoff}{2014}]%
        {leagueOfLegends}
\bibfield{author}{\bibinfo{person}{Todd Hoff}.}
  \bibinfo{year}{2014}\natexlab{}.
\newblock \bibinfo{booktitle}{\emph{How League Of Legends Scaled Chat To 70
  Million Players - It Takes Lots Of Minions}}.
\newblock
\urldef\tempurl%
\url{http://highscalability.com/blog/2014/10/13/how-league-of-legends-scaled-chat-to-70-million-players-it-t.html}
\showURL{%
\tempurl}


\bibitem[\protect\citeauthoryear{Holt, Bornholt, Zhang, Ports, Oskin, and
  Ceze}{Holt et~al\mbox{.}}{2016}]%
        {disciplined-inconsistency}
\bibfield{author}{\bibinfo{person}{Brandon Holt}, \bibinfo{person}{James
  Bornholt}, \bibinfo{person}{Irene Zhang}, \bibinfo{person}{Dan Ports},
  \bibinfo{person}{Mark Oskin}, {and} \bibinfo{person}{Luis Ceze}.}
  \bibinfo{year}{2016}\natexlab{}.
\newblock \showarticletitle{Disciplined Inconsistency with Consistency Types}.
  In \bibinfo{booktitle}{\emph{Proceedings of the Seventh ACM Symposium on
  Cloud Computing}}. \bibinfo{pages}{279–293}.
\newblock


\bibitem[\protect\citeauthoryear{Joshi}{Joshi}{2022}]%
        {joshiblog}
\bibfield{author}{\bibinfo{person}{Leena Joshi}.}
  \bibinfo{year}{2022}\natexlab{}.
\newblock \bibinfo{title}{How to simplify distributed app development with
  CRDTs}.
\newblock
\newblock
\newblock
\shownote{\url{https://techbeacon.com/app-dev-testing/how-simplify-distributed-app-development-crdts},
  retrieved May 31, 2022.}


\bibitem[\protect\citeauthoryear{Kaki, Priya, Sivaramakrishnan, and
  Jagannathan}{Kaki et~al\mbox{.}}{2019}]%
        {mrdts}
\bibfield{author}{\bibinfo{person}{Gowtham Kaki}, \bibinfo{person}{Swarn
  Priya}, \bibinfo{person}{KC Sivaramakrishnan}, {and} \bibinfo{person}{Suresh
  Jagannathan}.} \bibinfo{year}{2019}\natexlab{}.
\newblock \showarticletitle{Mergeable Replicated Data Types}.
\newblock \bibinfo{journal}{\emph{Proc. ACM Program. Lang.}}
  (\bibinfo{year}{2019}).
\newblock


\bibitem[\protect\citeauthoryear{Khamis, Ngo, Pichler, Suciu, and Wang}{Khamis
  et~al\mbox{.}}{2022}]%
        {khamisPODS22}
\bibfield{author}{\bibinfo{person}{Mahmoud~Abo Khamis},
  \bibinfo{person}{Hung~Q. Ngo}, \bibinfo{person}{Reinhard Pichler},
  \bibinfo{person}{Dan Suciu}, {and} \bibinfo{person}{Yisu~Remy Wang}.}
  \bibinfo{year}{2022}\natexlab{}.
\newblock \showarticletitle{Convergence of Datalog over (Pre-)Semirings}. In
  \bibinfo{booktitle}{\emph{PODS}}.
\newblock


\bibitem[\protect\citeauthoryear{Kleppmann}{Kleppmann}{2018}]%
        {kleppmannDatalog}
\bibfield{author}{\bibinfo{person}{Martin Kleppmann}.}
  \bibinfo{year}{2018}\natexlab{}.
\newblock \showarticletitle{Data structures as queries: Expressing CRDTs using
  Datalog}.
\newblock  (\bibinfo{year}{2018}).
\newblock
\urldef\tempurl%
\url{https://martin.kleppmann.com/2018/02/26/dagstuhl-data-consistency.html}
\showURL{%
\tempurl}


\bibitem[\protect\citeauthoryear{Kleppmann}{Kleppmann}{2019}]%
        {kleppmanColumnarCRDT}
\bibfield{author}{\bibinfo{person}{Martin Kleppmann}.}
  \bibinfo{year}{2019}\natexlab{}.
\newblock \showarticletitle{Experiment: columnar data encoding for Automerge}.
\newblock  (\bibinfo{year}{2019}).
\newblock
\urldef\tempurl%
\url{https://github.com/automerge/automerge-perf/blob/master/columnar/README.md}
\showURL{%
\tempurl}


\bibitem[\protect\citeauthoryear{Kleppmann}{Kleppmann}{2022}]%
        {automerge}
\bibfield{author}{\bibinfo{person}{Martin Kleppmann}.}
  \bibinfo{year}{2022}\natexlab{}.
\newblock \bibinfo{title}{Automerge}.
\newblock
\newblock
\urldef\tempurl%
\url{https://github.com/automerge/automerge}
\showURL{%
\tempurl}


\bibitem[\protect\citeauthoryear{Kleppmann and Beresford}{Kleppmann and
  Beresford}{2017}]%
        {jsonCrdt}
\bibfield{author}{\bibinfo{person}{Martin Kleppmann} {and}
  \bibinfo{person}{Alastair~R. Beresford}.} \bibinfo{year}{2017}\natexlab{}.
\newblock \showarticletitle{A Conflict-Free Replicated JSON Datatype}.
\newblock \bibinfo{journal}{\emph{IEEE Transactions on Parallel and Distributed
  Systems}} (\bibinfo{year}{2017}), \bibinfo{pages}{2733--2746}.
\newblock


\bibitem[\protect\citeauthoryear{Klophaus}{Klophaus}{2010}]%
        {riak}
\bibfield{author}{\bibinfo{person}{Rusty Klophaus}.}
  \bibinfo{year}{2010}\natexlab{}.
\newblock \showarticletitle{Riak Core: Building Distributed Applications
  without Shared State}. In \bibinfo{booktitle}{\emph{ACM SIGPLAN Commercial
  Users of Functional Programming}}.
\newblock


\bibitem[\protect\citeauthoryear{Kraska, Pang, Franklin, Madden, and
  Fekete}{Kraska et~al\mbox{.}}{2013}]%
        {mdcc}
\bibfield{author}{\bibinfo{person}{Tim Kraska}, \bibinfo{person}{Gene Pang},
  \bibinfo{person}{Michael~J Franklin}, \bibinfo{person}{Samuel Madden}, {and}
  \bibinfo{person}{Alan Fekete}.} \bibinfo{year}{2013}\natexlab{}.
\newblock \showarticletitle{MDCC: Multi-data center consistency}. In
  \bibinfo{booktitle}{\emph{Proceedings of the 8th ACM European Conference on
  Computer Systems}}. \bibinfo{pages}{113--126}.
\newblock


\bibitem[\protect\citeauthoryear{Kuper and Newton}{Kuper and Newton}{2013}]%
        {lvars}
\bibfield{author}{\bibinfo{person}{Lindsey Kuper} {and}
  \bibinfo{person}{Ryan~R. Newton}.} \bibinfo{year}{2013}\natexlab{}.
\newblock \showarticletitle{LVars: Lattice-Based Data Structures for
  Deterministic Parallelism}. In \bibinfo{booktitle}{\emph{Proceedings of the
  2nd ACM SIGPLAN Workshop on Functional High-Performance Computing}}.
  \bibinfo{pages}{71–84}.
\newblock


\bibitem[\protect\citeauthoryear{Laddad, Power, Milano, Cheung, and
  Hellerstein}{Laddad et~al\mbox{.}}{2022}]%
        {katara}
\bibfield{author}{\bibinfo{person}{Shadaj Laddad}, \bibinfo{person}{Conor
  Power}, \bibinfo{person}{Mae Milano}, \bibinfo{person}{Alvin Cheung}, {and}
  \bibinfo{person}{Joseph~M. Hellerstein}.} \bibinfo{year}{2022}\natexlab{}.
\newblock \bibinfo{title}{Katara: Synthesizing CRDTs with Verified Lifting}.
\newblock
\newblock
\urldef\tempurl%
\url{https://doi.org/10.48550/ARXIV.2205.12425}
\showDOI{\tempurl}


\bibitem[\protect\citeauthoryear{Lakshman and Malik}{Lakshman and
  Malik}{2010}]%
        {cassandra}
\bibfield{author}{\bibinfo{person}{Avinash Lakshman} {and}
  \bibinfo{person}{Prashant Malik}.} \bibinfo{year}{2010}\natexlab{}.
\newblock \showarticletitle{Cassandra: A Decentralized Structured Storage
  System}.
\newblock \bibinfo{journal}{\emph{SIGOPS Oper. Syst. Rev.}}
  (\bibinfo{year}{2010}), \bibinfo{pages}{35–40}.
\newblock


\bibitem[\protect\citeauthoryear{Lamport}{Lamport}{1978}]%
        {lamportTimestamps}
\bibfield{author}{\bibinfo{person}{Leslie Lamport}.}
  \bibinfo{year}{1978}\natexlab{}.
\newblock \showarticletitle{Time, Clocks, and the Ordering of Events in a
  Distributed System}.
\newblock \bibinfo{journal}{\emph{Commun. ACM}} (\bibinfo{year}{1978}),
  \bibinfo{pages}{558–565}.
\newblock


\bibitem[\protect\citeauthoryear{Li, Leit{\~a}o, Clement, Pregui{\c c}a,
  Rodrigues, and Vafeiadis}{Li et~al\mbox{.}}{2014}]%
        {redblue-automating}
\bibfield{author}{\bibinfo{person}{Cheng Li}, \bibinfo{person}{Joao
  Leit{\~a}o}, \bibinfo{person}{Allen Clement}, \bibinfo{person}{Nuno Pregui{\c
  c}a}, \bibinfo{person}{Rodrigo Rodrigues}, {and} \bibinfo{person}{Viktor
  Vafeiadis}.} \bibinfo{year}{2014}\natexlab{}.
\newblock \showarticletitle{Automating the Choice of Consistency Levels in
  Replicated Systems}. In \bibinfo{booktitle}{\emph{2014 USENIX Annual
  Technical Conference (USENIX ATC 14)}}. \bibinfo{pages}{281--292}.
\newblock


\bibitem[\protect\citeauthoryear{Li, Porto, Clement, Gehrke, Pregui{\c c}a, and
  Rodrigues}{Li et~al\mbox{.}}{2012}]%
        {redblue}
\bibfield{author}{\bibinfo{person}{Cheng Li}, \bibinfo{person}{Daniel Porto},
  \bibinfo{person}{Allen Clement}, \bibinfo{person}{Johannes Gehrke},
  \bibinfo{person}{Nuno Pregui{\c c}a}, {and} \bibinfo{person}{Rodrigo
  Rodrigues}.} \bibinfo{year}{2012}\natexlab{}.
\newblock \showarticletitle{Making {Geo-Replicated} Systems Fast as Possible,
  Consistent when Necessary}. In \bibinfo{booktitle}{\emph{10th USENIX
  Symposium on Operating Systems Design and Implementation (OSDI 12)}}.
\newblock


\bibitem[\protect\citeauthoryear{Litt, Lim, Kleppmann, and van Hardenberg}{Litt
  et~al\mbox{.}}{2021}]%
        {peritext}
\bibfield{author}{\bibinfo{person}{Geoffrey Litt}, \bibinfo{person}{Slim Lim},
  \bibinfo{person}{Martin Kleppmann}, {and} \bibinfo{person}{Peter van
  Hardenberg}.} \bibinfo{year}{2021}\natexlab{}.
\newblock \bibinfo{title}{Peritext: A CRDT for Rich-Text Collaboration}.
\newblock
\newblock
\urldef\tempurl%
\url{https://www.inkandswitch.com/peritext}
\showURL{%
\tempurl}


\bibitem[\protect\citeauthoryear{Liu, Magrino, Arden, George, and Myers}{Liu
  et~al\mbox{.}}{2014}]%
        {warranties}
\bibfield{author}{\bibinfo{person}{Jed Liu}, \bibinfo{person}{Tom Magrino},
  \bibinfo{person}{Owen Arden}, \bibinfo{person}{Michael~D George}, {and}
  \bibinfo{person}{Andrew~C Myers}.} \bibinfo{year}{2014}\natexlab{}.
\newblock \showarticletitle{Warranties for faster strong consistency}. In
  \bibinfo{booktitle}{\emph{11th USENIX Symposium on Networked Systems Design
  and Implementation (NSDI 14)}}. \bibinfo{pages}{503--517}.
\newblock


\bibitem[\protect\citeauthoryear{Lloyd, Freedman, Kaminsky, and Andersen}{Lloyd
  et~al\mbox{.}}{2011}]%
        {cops}
\bibfield{author}{\bibinfo{person}{Wyatt Lloyd}, \bibinfo{person}{Michael~J
  Freedman}, \bibinfo{person}{Michael Kaminsky}, {and} \bibinfo{person}{David~G
  Andersen}.} \bibinfo{year}{2011}\natexlab{}.
\newblock \showarticletitle{Don't settle for eventual: Scalable causal
  consistency for wide-area storage with COPS}. In
  \bibinfo{booktitle}{\emph{Proceedings of the Twenty-Third ACM Symposium on
  Operating Systems Principles}}. \bibinfo{pages}{401--416}.
\newblock


\bibitem[\protect\citeauthoryear{Martyanov}{Martyanov}{2018}]%
        {paypal_crdt}
\bibfield{author}{\bibinfo{person}{Dmitry Martyanov}.}
  \bibinfo{year}{2018}\natexlab{}.
\newblock \bibinfo{title}{CRDTs in Production}.
\newblock
\newblock
\urldef\tempurl%
\url{https://www.infoq.com/presentations/crdt-production}
\showURL{%
\tempurl}


\bibitem[\protect\citeauthoryear{Meiklejohn and Van~Roy}{Meiklejohn and
  Van~Roy}{2015}]%
        {lasp}
\bibfield{author}{\bibinfo{person}{Christopher Meiklejohn} {and}
  \bibinfo{person}{Peter Van~Roy}.} \bibinfo{year}{2015}\natexlab{}.
\newblock \showarticletitle{Lasp: A Language for Distributed, Coordination-Free
  Programming}. In \bibinfo{booktitle}{\emph{Proceedings of the 17th
  International Symposium on Principles and Practice of Declarative
  Programming}}. \bibinfo{pages}{184–195}.
\newblock


\bibitem[\protect\citeauthoryear{Milano and Myers}{Milano and Myers}{2018}]%
        {mixt}
\bibfield{author}{\bibinfo{person}{Mae Milano} {and} \bibinfo{person}{Andrew~C.
  Myers}.} \bibinfo{year}{2018}\natexlab{}.
\newblock \showarticletitle{MixT: A Language for Mixing Consistency in
  Geodistributed Transactions}. In \bibinfo{booktitle}{\emph{Proceedings of the
  39th ACM SIGPLAN Conference on Programming Language Design and
  Implementation}}. \bibinfo{pages}{226–241}.
\newblock


\bibitem[\protect\citeauthoryear{Milano, Recto, Magrino, and Myers}{Milano
  et~al\mbox{.}}{2019}]%
        {gallifrey}
\bibfield{author}{\bibinfo{person}{Mae Milano}, \bibinfo{person}{Rolph Recto},
  \bibinfo{person}{Tom Magrino}, {and} \bibinfo{person}{Andrew~C. Myers}.}
  \bibinfo{year}{2019}\natexlab{}.
\newblock \showarticletitle{{A Tour of Gallifrey, a Language for Geodistributed
  Programming}}. In \bibinfo{booktitle}{\emph{3rd Summit on Advances in
  Programming Languages (SNAPL 2019)}}. \bibinfo{pages}{11:1--11:19}.
\newblock


\bibitem[\protect\citeauthoryear{Neely and Rusu}{Neely and Rusu}{2022}]%
        {rust_crdt}
\bibfield{author}{\bibinfo{person}{Tyler Neely} {and} \bibinfo{person}{David
  Rusu}.} \bibinfo{year}{2022}\natexlab{}.
\newblock \bibinfo{title}{crdts: family of thoroughly tested hybrid crdt's}.
\newblock
\newblock
\urldef\tempurl%
\url{https://github.com/rust-crdt/rust-crdt}
\showURL{%
\tempurl}


\bibitem[\protect\citeauthoryear{Nicolaescu, Jahns, Derntl, and
  Klamma}{Nicolaescu et~al\mbox{.}}{2015}]%
        {yjs}
\bibfield{author}{\bibinfo{person}{Petru Nicolaescu}, \bibinfo{person}{Kevin
  Jahns}, \bibinfo{person}{Michael Derntl}, {and} \bibinfo{person}{Ralf
  Klamma}.} \bibinfo{year}{2015}\natexlab{}.
\newblock \showarticletitle{Yjs: A Framework for Near Real-Time P2P Shared
  Editing on Arbitrary Data Types}. In \bibinfo{booktitle}{\emph{Proceedings of
  the 15th International Conference on Engineering the Web in the Big Data Era
  - Volume 9114}}. \bibinfo{pages}{675–678}.
\newblock


\bibitem[\protect\citeauthoryear{O'Neil}{O'Neil}{1986}]%
        {escrow}
\bibfield{author}{\bibinfo{person}{Patrick~E O'Neil}.}
  \bibinfo{year}{1986}\natexlab{}.
\newblock \showarticletitle{The escrow transactional method}.
\newblock \bibinfo{journal}{\emph{ACM Transactions on Database Systems (TODS)}}
  (\bibinfo{year}{1986}), \bibinfo{pages}{405--430}.
\newblock


\bibitem[\protect\citeauthoryear{Pang, C\'{a}ceres, Burrows, Chen, Dave,
  Germer, Golynski, Graney, Kang, Kissner, Korn, Parmar, Richards, and
  Wang}{Pang et~al\mbox{.}}{2019}]%
        {zanzibar}
\bibfield{author}{\bibinfo{person}{Ruoming Pang}, \bibinfo{person}{Ram\'{o}n
  C\'{a}ceres}, \bibinfo{person}{Mike Burrows}, \bibinfo{person}{Zhifeng Chen},
  \bibinfo{person}{Pratik Dave}, \bibinfo{person}{Nathan Germer},
  \bibinfo{person}{Alexander Golynski}, \bibinfo{person}{Kevin Graney},
  \bibinfo{person}{Nina Kang}, \bibinfo{person}{Lea Kissner},
  \bibinfo{person}{Jeffrey~L. Korn}, \bibinfo{person}{Abhishek Parmar},
  \bibinfo{person}{Christopher~D. Richards}, {and} \bibinfo{person}{Mengzhi
  Wang}.} \bibinfo{year}{2019}\natexlab{}.
\newblock \showarticletitle{Zanzibar: Google's Consistent, Global Authorization
  System}. In \bibinfo{booktitle}{\emph{Proceedings of the 2019 USENIX
  Conference on Usenix Annual Technical Conference}} (Renton, WA, USA)
  \emph{(\bibinfo{series}{USENIX ATC '19})}. \bibinfo{publisher}{USENIX
  Association}, \bibinfo{address}{USA}, \bibinfo{pages}{33–46}.
\newblock
\showISBNx{9781939133038}


\bibitem[\protect\citeauthoryear{Ramamritham and Pu}{Ramamritham and
  Pu}{1995}]%
        {epsilonSerializability}
\bibfield{author}{\bibinfo{person}{Krithi Ramamritham} {and}
  \bibinfo{person}{Calton Pu}.} \bibinfo{year}{1995}\natexlab{}.
\newblock \showarticletitle{A formal characterization of epsilon
  serializability}.
\newblock \bibinfo{journal}{\emph{IEEE Transactions on Knowledge and Data
  Engineering}} (\bibinfo{year}{1995}), \bibinfo{pages}{997--1007}.
\newblock


\bibitem[\protect\citeauthoryear{Rinberg, Solomon, Shlomo, Khazma, Lushi,
  Keidar, and Ta-Shma}{Rinberg et~al\mbox{.}}{2022}]%
        {dson}
\bibfield{author}{\bibinfo{person}{Arik Rinberg}, \bibinfo{person}{Tomer
  Solomon}, \bibinfo{person}{Roee Shlomo}, \bibinfo{person}{Guy Khazma},
  \bibinfo{person}{Gal Lushi}, \bibinfo{person}{Idit Keidar}, {and}
  \bibinfo{person}{Paula Ta-Shma}.} \bibinfo{year}{2022}\natexlab{}.
\newblock \showarticletitle{DSON: JSON CRDT using delta-mutations for document
  stores}.
\newblock \bibinfo{journal}{\emph{Proceedings of the VLDB Endowment}}
  (\bibinfo{year}{2022}), \bibinfo{pages}{1053--1065}.
\newblock


\bibitem[\protect\citeauthoryear{Roestenburg, Williams, and Bakker}{Roestenburg
  et~al\mbox{.}}{2016}]%
        {akka}
\bibfield{author}{\bibinfo{person}{Raymond Roestenburg}, \bibinfo{person}{Rob
  Williams}, {and} \bibinfo{person}{Robertus Bakker}.}
  \bibinfo{year}{2016}\natexlab{}.
\newblock \bibinfo{booktitle}{\emph{Akka in action}}.
\newblock


\bibitem[\protect\citeauthoryear{Rofls}{Rofls}{2022}]%
        {xiBlog}
\bibfield{author}{\bibinfo{person}{Colin Rofls}.}
  \bibinfo{year}{2022}\natexlab{}.
\newblock \bibinfo{title}{CRDT - The Xi Text Engine}.
\newblock
\newblock
\newblock
\shownote{\url{https://xi-editor.io/docs/crdt-details.html/}, retrieved June 1,
  2022.}


\bibitem[\protect\citeauthoryear{Roy, Kot, Bender, Ding, Hojjat, Koch, Foster,
  and Gehrke}{Roy et~al\mbox{.}}{2015}]%
        {homeostasis}
\bibfield{author}{\bibinfo{person}{Sudip Roy}, \bibinfo{person}{Lucja Kot},
  \bibinfo{person}{Gabriel Bender}, \bibinfo{person}{Bailu Ding},
  \bibinfo{person}{Hossein Hojjat}, \bibinfo{person}{Christoph Koch},
  \bibinfo{person}{Nate Foster}, {and} \bibinfo{person}{Johannes Gehrke}.}
  \bibinfo{year}{2015}\natexlab{}.
\newblock \showarticletitle{The homeostasis protocol: Avoiding transaction
  coordination through program analysis}. In
  \bibinfo{booktitle}{\emph{Proceedings of the 2015 ACM SIGMOD International
  Conference on Management of Data}}. \bibinfo{pages}{1311--1326}.
\newblock


\bibitem[\protect\citeauthoryear{Shapiro, Pregui{\c{c}}a, Baquero, and
  Zawirski}{Shapiro et~al\mbox{.}}{2011a}]%
        {shapiro2011comprehensive}
\bibfield{author}{\bibinfo{person}{Marc Shapiro}, \bibinfo{person}{Nuno
  Pregui{\c{c}}a}, \bibinfo{person}{Carlos Baquero}, {and}
  \bibinfo{person}{Marek Zawirski}.} \bibinfo{year}{2011}\natexlab{a}.
\newblock \emph{\bibinfo{title}{A comprehensive study of convergent and
  commutative replicated data types}}.
\newblock \bibinfo{thesistype}{Ph.D. Dissertation}.
  \bibinfo{school}{Inria--Centre Paris-Rocquencourt; INRIA}.
\newblock


\bibitem[\protect\citeauthoryear{Shapiro, Pregui{\c{c}}a, Baquero, and
  Zawirski}{Shapiro et~al\mbox{.}}{2011b}]%
        {CRDTs}
\bibfield{author}{\bibinfo{person}{Marc Shapiro}, \bibinfo{person}{Nuno
  Pregui{\c{c}}a}, \bibinfo{person}{Carlos Baquero}, {and}
  \bibinfo{person}{Marek Zawirski}.} \bibinfo{year}{2011}\natexlab{b}.
\newblock \showarticletitle{Conflict-free replicated data types}. In
  \bibinfo{booktitle}{\emph{Symposium on Self-Stabilizing Systems}}.
  \bibinfo{pages}{386--400}.
\newblock


\bibitem[\protect\citeauthoryear{Shi, Pruett, Doherty, Han, Petrov, Carrig,
  Hugg, and Bronson}{Shi et~al\mbox{.}}{2020}]%
        {flighttracker}
\bibfield{author}{\bibinfo{person}{Xiao Shi}, \bibinfo{person}{Scott Pruett},
  \bibinfo{person}{Kevin Doherty}, \bibinfo{person}{Jinyu Han},
  \bibinfo{person}{Dmitri Petrov}, \bibinfo{person}{Jim Carrig},
  \bibinfo{person}{John Hugg}, {and} \bibinfo{person}{Nathan Bronson}.}
  \bibinfo{year}{2020}\natexlab{}.
\newblock \showarticletitle{$\{$FlightTracker$\}$: Consistency across
  $\{$Read-Optimized$\}$ Online Stores at Facebook}. In
  \bibinfo{booktitle}{\emph{14th USENIX Symposium on Operating Systems Design
  and Implementation (OSDI 20)}}. \bibinfo{pages}{407--423}.
\newblock


\bibitem[\protect\citeauthoryear{Sivaramakrishnan, Kaki, and
  Jagannathan}{Sivaramakrishnan et~al\mbox{.}}{2015}]%
        {quelea}
\bibfield{author}{\bibinfo{person}{KC Sivaramakrishnan},
  \bibinfo{person}{Gowtham Kaki}, {and} \bibinfo{person}{Suresh Jagannathan}.}
  \bibinfo{year}{2015}\natexlab{}.
\newblock \showarticletitle{Declarative Programming over Eventually Consistent
  Data Stores}. In \bibinfo{booktitle}{\emph{Proceedings of the 36th ACM
  SIGPLAN Conference on Programming Language Design and Implementation}}.
  \bibinfo{pages}{413–424}.
\newblock


\bibitem[\protect\citeauthoryear{Sreekanti, Wu, Lin, Schleier-Smith, Gonzalez,
  Hellerstein, and Tumanov}{Sreekanti et~al\mbox{.}}{2020}]%
        {cloudburst}
\bibfield{author}{\bibinfo{person}{Vikram Sreekanti},
  \bibinfo{person}{Chenggang Wu}, \bibinfo{person}{Xiayue~Charles Lin},
  \bibinfo{person}{Johann Schleier-Smith}, \bibinfo{person}{Joseph~E.
  Gonzalez}, \bibinfo{person}{Joseph~M. Hellerstein}, {and}
  \bibinfo{person}{Alexey Tumanov}.} \bibinfo{year}{2020}\natexlab{}.
\newblock \showarticletitle{Cloudburst: Stateful Functions-as-a-Service}.
\newblock \bibinfo{journal}{\emph{Proc. VLDB Endow.}} \bibinfo{volume}{13},
  \bibinfo{number}{12} (\bibinfo{date}{jul} \bibinfo{year}{2020}),
  \bibinfo{pages}{2438–2452}.
\newblock
\showISSN{2150-8097}
\urldef\tempurl%
\url{https://doi.org/10.14778/3407790.3407836}
\showDOI{\tempurl}


\bibitem[\protect\citeauthoryear{Stonebraker and Rowe}{Stonebraker and
  Rowe}{1986}]%
        {postgres}
\bibfield{author}{\bibinfo{person}{Michael Stonebraker} {and}
  \bibinfo{person}{Lawrence~A. Rowe}.} \bibinfo{year}{1986}\natexlab{}.
\newblock \showarticletitle{The Design of POSTGRES}. In
  \bibinfo{booktitle}{\emph{Proceedings of the 1986 ACM SIGMOD International
  Conference on Management of Data}} (Washington, D.C., USA)
  \emph{(\bibinfo{series}{SIGMOD '86})}. \bibinfo{publisher}{Association for
  Computing Machinery}, \bibinfo{address}{New York, NY, USA},
  \bibinfo{pages}{340–355}.
\newblock
\showISBNx{0897911911}
\urldef\tempurl%
\url{https://doi.org/10.1145/16894.16888}
\showDOI{\tempurl}


\bibitem[\protect\citeauthoryear{Sypytkowski}{Sypytkowski}{2022}]%
        {bartoszBlog}
\bibfield{author}{\bibinfo{person}{Bartosz Sypytkowski}.}
  \bibinfo{year}{2022}\natexlab{}.
\newblock \bibinfo{title}{An Introduction to State-based CRDTs}.
\newblock
\newblock
\newblock
\shownote{\url{https://bartoszsypytkowski.com/the-state-of-a-state-based-crdts/},
  retrieved June 1, 2022.}


\bibitem[\protect\citeauthoryear{Terry, Prabhakaran, Kotla, Balakrishnan,
  Aguilera, and Abu-Libdeh}{Terry et~al\mbox{.}}{2013}]%
        {pileus}
\bibfield{author}{\bibinfo{person}{Douglas~B Terry}, \bibinfo{person}{Vijayan
  Prabhakaran}, \bibinfo{person}{Ramakrishna Kotla}, \bibinfo{person}{Mahesh
  Balakrishnan}, \bibinfo{person}{Marcos~K Aguilera}, {and}
  \bibinfo{person}{Hussam Abu-Libdeh}.} \bibinfo{year}{2013}\natexlab{}.
\newblock \showarticletitle{Consistency-based service level agreements for
  cloud storage}. In \bibinfo{booktitle}{\emph{Proceedings of the Twenty-Fourth
  ACM Symposium on Operating Systems Principles}}. \bibinfo{pages}{309--324}.
\newblock


\bibitem[\protect\citeauthoryear{Wang, Enea, Mutluergil, and Petri}{Wang
  et~al\mbox{.}}{2019}]%
        {ralinearizability}
\bibfield{author}{\bibinfo{person}{Chao Wang}, \bibinfo{person}{Constantin
  Enea}, \bibinfo{person}{Suha~Orhun Mutluergil}, {and}
  \bibinfo{person}{Gustavo Petri}.} \bibinfo{year}{2019}\natexlab{}.
\newblock \showarticletitle{Replication-Aware Linearizability}. In
  \bibinfo{booktitle}{\emph{Proceedings of the 40th ACM SIGPLAN Conference on
  Programming Language Design and Implementation}}. \bibinfo{pages}{980–993}.
\newblock


\bibitem[\protect\citeauthoryear{Weiss, Urso, and Molli}{Weiss
  et~al\mbox{.}}{2007}]%
        {wooki}
\bibfield{author}{\bibinfo{person}{St{\'e}phane Weiss}, \bibinfo{person}{Pascal
  Urso}, {and} \bibinfo{person}{Pascal Molli}.}
  \bibinfo{year}{2007}\natexlab{}.
\newblock \showarticletitle{{Wooki}: A {P2P} Wiki-Based Collaborative Writing
  Tool}. In \bibinfo{booktitle}{\emph{8th International Conference on Web
  Information Systems Engineering}}. \bibinfo{pages}{503--512}.
\newblock


\bibitem[\protect\citeauthoryear{Weiss, Urso, and Molli}{Weiss
  et~al\mbox{.}}{2009}]%
        {logoot}
\bibfield{author}{\bibinfo{person}{Stephane Weiss}, \bibinfo{person}{Pascal
  Urso}, {and} \bibinfo{person}{Pascal Molli}.}
  \bibinfo{year}{2009}\natexlab{}.
\newblock \showarticletitle{Logoot: A Scalable Optimistic Replication Algorithm
  for Collaborative Editing on P2P Networks}. In \bibinfo{booktitle}{\emph{2009
  29th IEEE International Conference on Distributed Computing Systems}}.
  \bibinfo{pages}{404--412}.
\newblock


\bibitem[\protect\citeauthoryear{Wu, Faleiro, Lin, and Hellerstein}{Wu
  et~al\mbox{.}}{2018}]%
        {anna}
\bibfield{author}{\bibinfo{person}{Chenggang Wu}, \bibinfo{person}{Jose
  Faleiro}, \bibinfo{person}{Yihan Lin}, {and} \bibinfo{person}{Joseph
  Hellerstein}.} \bibinfo{year}{2018}\natexlab{}.
\newblock \showarticletitle{Anna: A KVS for Any Scale}. In
  \bibinfo{booktitle}{\emph{2018 IEEE 34th International Conference on Data
  Engineering (ICDE)}}. \bibinfo{pages}{401--412}.
\newblock


\bibitem[\protect\citeauthoryear{Zawirski, Pregui{\c{c}}a, Duarte, Bieniusa,
  Balegas, and Shapiro}{Zawirski et~al\mbox{.}}{2015}]%
        {zawirski2015write}
\bibfield{author}{\bibinfo{person}{Marek Zawirski}, \bibinfo{person}{Nuno
  Pregui{\c{c}}a}, \bibinfo{person}{S{\'e}rgio Duarte},
  \bibinfo{person}{Annette Bieniusa}, \bibinfo{person}{Valter Balegas}, {and}
  \bibinfo{person}{Marc Shapiro}.} \bibinfo{year}{2015}\natexlab{}.
\newblock \showarticletitle{Write fast, read in the past: Causal consistency
  for client-side applications}. In \bibinfo{booktitle}{\emph{Proceedings of
  the 16th Annual Middleware Conference}}. \bibinfo{pages}{75--87}.
\newblock


\bibitem[\protect\citeauthoryear{Zhao and Haller}{Zhao and Haller}{2018}]%
        {observable-atomic-consistency}
\bibfield{author}{\bibinfo{person}{Xin Zhao} {and} \bibinfo{person}{Philipp
  Haller}.} \bibinfo{year}{2018}\natexlab{}.
\newblock \showarticletitle{Observable Atomic Consistency for CvRDTs}. In
  \bibinfo{booktitle}{\emph{Proceedings of the 8th ACM SIGPLAN International
  Workshop on Programming Based on Actors, Agents, and Decentralized Control}}.
  \bibinfo{pages}{23–32}.
\newblock


\end{thebibliography}

\end{document}